\documentclass[12pt]{revtex4-2}
\usepackage[colorlinks]{hyperref} 
\usepackage{amsmath}
\usepackage{amsfonts}

\def\exp{\hbox{\rm exp}}
\smallskip 
\def\<{{\langle }}
\def\>{{\rangle }}

\def\exp{\hbox{\rm exp}} 
\smallskip
\def\<{{\langle }}
\def\>{{\rangle }}

\usepackage{color}
\usepackage{graphicx}%
\usepackage{dcolumn}%
\usepackage{bm}%
\usepackage{hyperref}%
\usepackage{tikz}

\begin{document} 

\title[Hidden Inverse in Superconducting processors]{Quantum error mitigation by hidden inverses protocol in superconducting quantum devices}
\thanks{This manuscript has been authored by UT-Battelle, LLC, under Contract No. DE-AC0500OR22725 with the U.S. Department of Energy. The United States Government retains and the publisher, by accepting the article for publication, acknowledges that the United States Government retains a non-exclusive, paid-up, irrevocable, world-wide license to publish or reproduce the published form of this manuscript, or allow others to do so, for the United States Government purposes. The Department of Energy will provide public access to these results of federally sponsored research in accordance with the DOE Public Access Plan.}%

\author{V Leyton-Ortega$^1$, S Majumder$^{2, 3}$, R C Pooser$^1$}
\affiliation{$^1$Computational Sciences and Engineering Division, Oak Ridge National Laboratory, Oak Ridge, TN 37831, USA}
\affiliation{$^2$Duke Quantum Center, Duke University, Durham, NC 27701, USA}
\affiliation{$^3$Department of Electrical and Computer Engineering, Duke University, Durham, NC 27708, USA}
\email{pooserrc@ornl.gov}

\date{\today}

\begin{abstract}    We present a method to improve the convergence of variational algorithms based on hidden inverses to mitigate coherent errors. In the context of error mitigation, this means replacing the on hardware implementation of certain Hermitian gates with their inverses. Doing so results in noise cancellation and a more resilient quantum circuit. This approach improves performance in a variety of two-qubit error models where the noise operator also inverts with the gate inversion. We apply the mitigation scheme on superconducting quantum processors running the variational quantum eigensolver (VQE) algorithm to find the H$_{\rm 2}$ ground-state energy. When implemented on superconducting hardware we find that the mitigation scheme effectively reduces the energy fluctuations in the parameter learning path in VQE, reducing the number of iterations for a converged value. We also provide a detailed numerical simulation of VQE performance under different noise models and explore how hidden inverses \& randomized compiling affect the underlying loss landscape of the learning problem. These simulations help explain our experimental hardware outcomes, helping to connect lower-level gate performance to application-specific behavior in contrast to metrics like fidelity which often do not provide an intuitive insight into observed high level performance.   
\end{abstract}


\maketitle 
\section{Introduction} \label{sec:intro}

A significant ingredient to improve algorithmic performance and potential quantum advantage in noisy-intermediate-scale quantum (NISQ) computers is the development of reliable quantum error mitigation (QEM) schemes.
A QEM protocol, combined with low-noise measurements or noise-mitigated measurements, allows one to detect and correct for errors during execution of a circuit or in post-processing. Some QEM techniques share similarities with quantum control protocols designed to reduce noise (such as dynamical decoupling) on the path to fault tolerant operations.  

Error mitigation techniques are useful for the NISQ period before fault tolerant quantum error correction (QEC) is fully scalable, although recent demonstrations provide great promise for the future~\cite{corcoles_demonstration_2015,tzitrin_fault-tolerant_2021,hilder_fault-tolerant_2021,krinner_realizing_2021,egan_fault-tolerant_2021,ryan-anderson_realization_2021}.
Error mitigation techniques usually aim to reduce the effects of noise in state preparation and measurement. For example, Richardson extrapolation is used to estimate expectation values in the noise-free limit by effectively amplifying noise via gate repetitions or gate pulse stretching~\cite{li2017efficient,PhysRevLett.119.180509}. Quasiprobability decomposition estimates the noise-free limit via repeated measurements~\cite{otten2019recovering,endo2018practical}. Experimental implementations of QEM can show results close to chemical accuracy for the ground state energy for molecules such as H${}_2$ and LiH using the variational quantum eigensolver method (VQE)~\cite{kandala2019error}. Depending on the algorithm or application, a variety of QEM techniques can be combined:  quantum subspace expansion~\cite{PhysRevA.95.042308};  symmetry verification that designs ans\"atze with unitary components that conserve the particle number and spin symmetries~\cite{mcardle2019error, bonet2018low}; ``error reduction'' which executes QEC circuits on on a single qubit and uses post-processing to mitigate errors~\cite{otten2019accounting};
readout error mitigation~\cite{maciejewski2020mitigation,chen2019detector, kwon2020hybrid};
machine learning techniques based on approximate error models~\cite{strikis2020learning,czarnik2020error,zlokapa2020deep}; and stochastic QEM that estimates errors based on continuous analog quantum simulation \cite{sun2020practical}. 

Hidden inverses do not require additional quantum resources in terms of qubits or extra observable measurements. 
The method shares some characteristics in common with quasi-probability methods~\cite{PhysRevLett.119.180509,endo2018practical},
which aim to cancel out noise processes by probabilistically implementing the inverse process as a compiled set. 
On the other hand, HI replaces certain gates that already exist in the circuit with their inverse, without further modifications to the ans\"atz. 

Below we review the HI model followed by demonstrations of its utility on superconducting hardware and simulations which verify the experimental overlap with our model.  In Section \ref{sec:gateinv}, we introduce gate inversion at the pulse level in  superconducting devices, followed by gate verification and noise boosting by quantum process tomography and unitary folding experiments. Section \ref{sec:vqe} applies the HI to VQE experiments to find the ground state energy of H$_2$, and we compare its effectiveness with randomized compilation. In section \ref{sec:dis} we finish with discuss our results and numerical simulations and conclusions in Sections \ref{sec:dis} and \ref{sec:conc}, respectively. 


{\it Hidden inverses model:} We can consider the \emph{on-hardware} quantum gate $\tilde{ U}$ as the ideal quantum gate $U$ (unitary quantum operation) transformed by a coherent error map $E$ as $\tilde{U} = U \cdot E$ (the order $E \cdot U$ is valid as well). Therefore, in this model, the output of an ideal quantum circuit $U_{\rm QC} = U_N \cdot U_{N-1} \cdots U_1$ is biased by the coherent error injected on each gate $U_i$ as $| \psi_{\rm out} \rangle  =  (U_N \cdot E_N)\cdots (U_1 \cdot E_1) |\psi_{\rm in}\rangle$, with input $|\psi_{\rm in} \rangle$. Certain widely used gates in quantum computing, such as $CX$ and $H$, are self-inverses. In principle, this Hermitian character of such gates $U_i$ in $U_{\rm QC}$ makes the quantum gate invariant under the gate inversion $U_i \rightarrow U_i^{-1}$. However, the gate inversion on hardware produces a different operation with beneficial consequences in reducing coherent noise effects. For instance, in the trivial case of $U_{\rm QC}= (U_0 \cdot E_0 ) \cdot \tilde{U}_1 \cdot (U_0 \cdot E_0 )$, replacing the third gate on the r.h.s. with its inverse, the gate $(U_0 \cdot E_0 )^{-1} = E^{-1}_0 \cdot U_0$, will produce the quantum circuit $U'_{\rm QC} = U_0 \cdot \tilde{U}_1 \cdot U_0 + O([\tilde{U}_1, E_0^{-1}])$ that essentially is less affected by the coherent noise and then with a better output in terms of fidelity for some $\tilde{U}_1$. This central idea of the HI is to mitigate the coherent noise by replacing some operations with their inverses in quantum circuits. The HI is introduced in \cite{zhang2021} as a tool for reducing the effect coherent error in trapped ion-based architecture, where induced over-rotations and phase misalignment for $CX$ gates were efficiently mitigated. Hidden inverses were also shown to improve performance of VQE in simulation under ion trap noise models~\cite{Kubra2021AQT}. In this work, we demonstrate an application of the HI in superconducting-based architecture, where where the pulse level of compilation must be used in order to construct custom inverse entangling gates. We verify the similarity between the default gate and its inverted version by comparing the real parts of their quantum process matrices. In addition, we measure the effect of the HI in a unitary folding on-hardware experiment. As a final test, we consider the effects of the HI in the learning path of a variational quantum eigenstate task, and we compare its efficiency with randomized compilations widely used for coherent noise mitigation. We perform a detailed numerical simulation of how various types of noise affect the performance of a VQE problem. We also study the effect of hidden inverses and randomized compiling on the loss landscape underlying the optimization problem.

\section{Gate inversion on superconducting based architectures} \label{sec:gateinv}

\begin{figure}
    \centering
    \includegraphics[width=\textwidth]{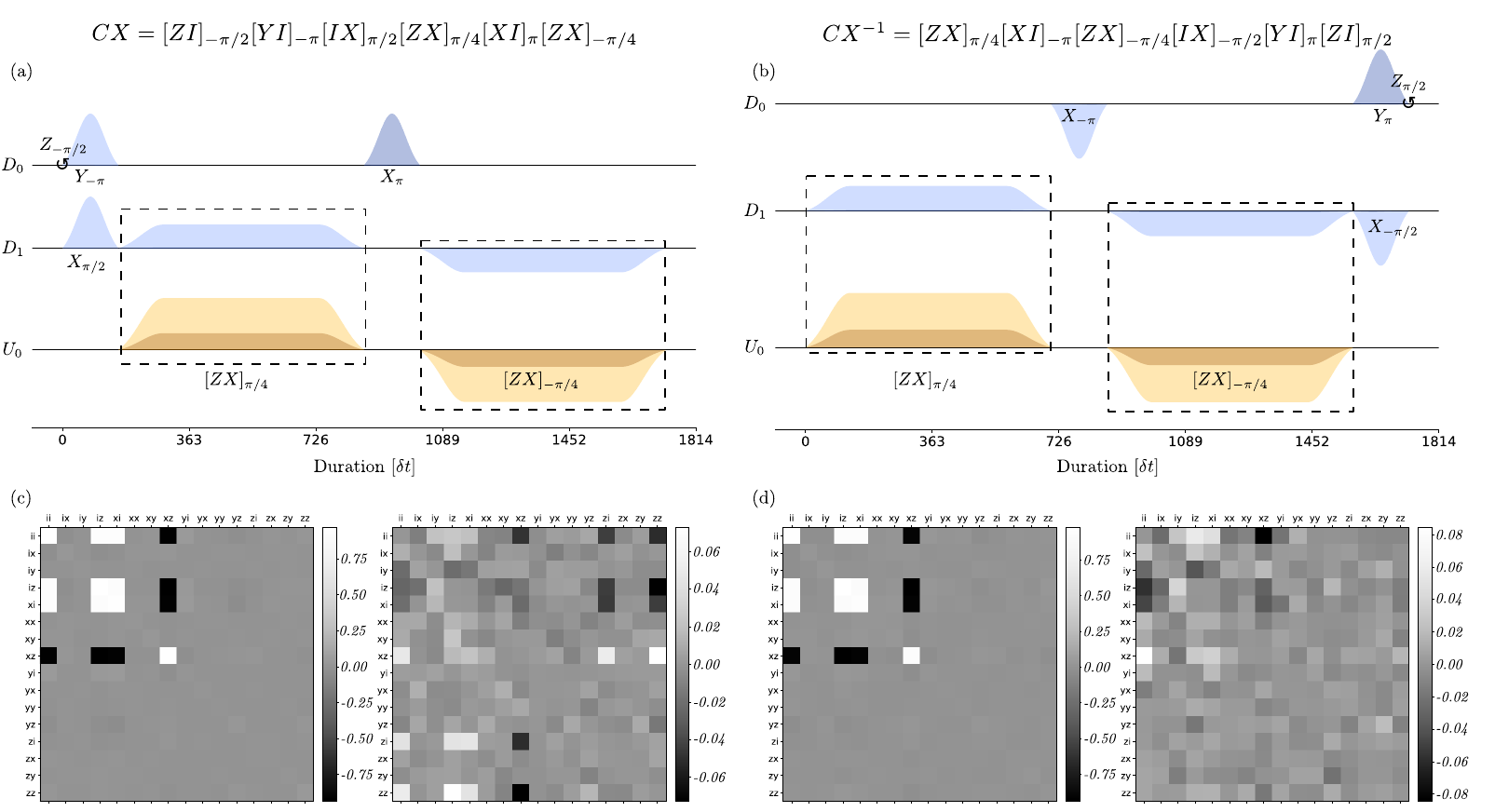}
    \caption{{\it Pulse profile for $CX^{-1}$ and verification:} We implemented a procedure to transform the pulse profile of a given gate, in this case, the $CX$ gate for the qubit pair [0,1] in the IBMQ-Montreal backend, presented in panel (a) into a pulse profile that represents its inverse, presented in panel (b). In addition, we have written the resulting compilation. In panels (c) and (d), we present the process matrix that verifies the operational equivalence between $CX$ and $CX^{-1}$.}
    \label{fig:cxcxinv}
\end{figure}

In most of the superconducting architectures, the $CX$ gate is built from an approximated cross-resonance quantum operation $CR = [ZX]_{\pi/2} = \exp(- i \pi \sigma_1^z \sigma_2^x / 4)$, introduced in \cite{Rigetti2010}. The $CR$ level of approximation defines the quality of $CX$ that can be used on hardware. 
Thus, it is imperative to find mechanisms to mitigate the level of coherent noise in this gate. The $CX$ gate is locally equivalent to $CR$ by $CX = [ZI]_{-\pi/2} [IX]_{-\pi/2} CR$. A better compilation of $CX$ is used in the IBM quantum device by echoing the $CR$ to mitigate the effects of undesired Hamiltonian terms \cite{Sheldon2021}. 
Every operation described in the compilation structure represents pulse construction and delivery instructions. We modify these instructions to mimic the inverted gate. In order to illustrate this procedure, we consider a simple example of the $\pi$-pulse construction in appendix A. We can generally define the pulse profile throughout the set of amplitudes $\lbrace u_0, \cdots, u_N \rbrace$ (waveform) or instructions according to the qiskit library of default functions (parametrized). 
We extract this information from default gates, and we define a new set of amplitudes $\lbrace -u_N, -u_{N-1}, \cdots, -u_0 \rbrace$ or new instructions for the parameterized version.
In Fig.\ref{fig:cxcxinv} (a) and (b), we present the default pulse and its inverse, respectively, for $CX_{01}$ on the IBMQ-Montreal backend. 
In principle, because the gate is self-inverse, the new gate should represent the same quantum process without adding additional noise. 
We verify this assumption by performing quantum process tomography to the gates $CX$ and $CX^{-1}$ and comparing their process matrix ($\chi$-matrix); see Fig. \ref{fig:cxcxinv} (c) and (d). 
The real part of the $\chi$-matrix for $CX^{-1}$ and $CX$ are similar, indicating that these gates represent the same quantum process (up to noise effects). 
On the other hand, their imaginary parts are different and somewhat statistically relevant (${\rm max} (\chi) > 1/\sqrt{N_S} \sim 10^{-2}$, $N_S = 5000$ the number of shots), which indicates slightly different noise properties. 
Therefore, the gates represent a similar, albeit not exactly identical, quantum process; it remains to test if the application of the new gate can be used to mitigate noise inside an algorithm on superconducting hardware. 

\begin{figure}
    \centering
    \includegraphics[width=\textwidth]{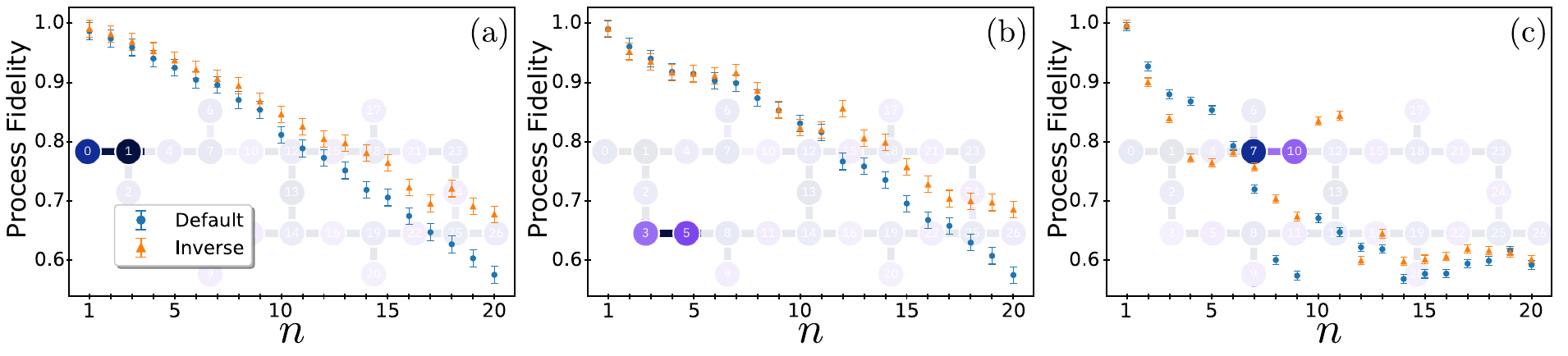}
    \caption{{\it Effective noise scaling on $CX$ gate - IBMQ-Montreal:} This plot presents the unitary folding test in two configurations, using the default gate ${\cal U} = CX \ CX$ (blue circles) and the inverse ${\cal U} = CX^{-1} CX$ (orange triangles). Additionally, we consider different backend sectors, showing the noise scaling at different base noise levels. We explored different setups for different values of error rates of local (circle darkness) and nonlocal (line darkness) operations; there, the darker the color of the region in the backend representation, the better is the performance of the quantum operation in terms of fidelity.} 
    \label{fig:foldingExp}
\end{figure}

{\it Unitary folding test:} Next, we consider a slightly different version of the noise scaling introduced \cite{Tiron2020} to examine the inverse gate's dependence on noise injection. 
In our version, we replace $CX$ with $CX \, {\cal U}^n $, where $\cal U$, in theory, is equivalent to the identity operator and $n$ is a positive number. 
In principle, this replacement should not alter the quantum operation defined by $CX$ and has no logical effect. 
Now, we choose two versions for $\cal U$, a first version ${\cal U} = CX \, CX$ to scale the effective noise on $CX$, and a second version where ${\cal U} = CX^{-1} CX$, which tests the mitigation on $CX$ by applying its inverse. 
Figure \ref{fig:foldingExp} shows how the effective noise on $CX$ scales in different sectors of the IBMQ-Montreal backend. 
We chose three sectors with different one- and two-qubit operational fidelity. 
Notably, in two of the sectors, noise mitigation is evident by the increase in fidelity as the number of gate repetitions increases. 

\section{Application to Variational Quantum algorithms} \label{sec:vqe}

\begin{widetext}
    \begin{figure}
        \centering
        \includegraphics[width =
        0.8\textwidth]{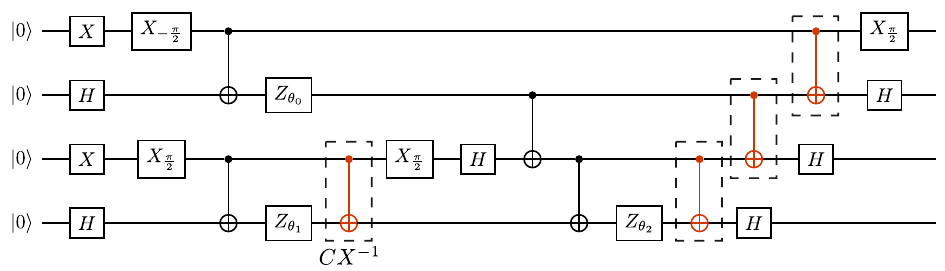}
        \caption{{\it The hidden inverse in the UCC3 ansatz:} The particular arrangement of the $CX$ gates in the UCC3 ansatz allows us to apply their inverse (gates inside the dashed squares) in front of default ones to maximize the coherent noise mitigation.}
        \label{fig:ucc3}
    \end{figure}
\end{widetext}

As a representative algorithm, we consider finding the ground state energy of the $H_2$ molecule using VQE. This parametrized algorithm updates quantum circuit's rotational parameters to minimize the Hamiltonian's expectation value $E_\theta$ until it converges. We consider the 3-parameter approximate unitary coupled-cluster ansatz for four qubits and 2 electrons (referred to as ``UCC3'' in \cite{McCaskey2019Benchmark}), see Fig.~\ref{fig:ucc3}. 
We systematically invert some of the $CX$ gates in the UCC3 ansatz to cancel the coherent noise present in CX. After every $CX$ applied on qubits $(i,j)$, we replace the next $CX$ applied on the same pair by its inverse. In Fig.~\ref{fig:ucc3}, we indicate by dashed squares the position of $CX^{-1}$. For the optimization part of the algorithm, we consider a gradient-free and gradient-based solvers to explore the influence of coherent error mitigation in the learning process under these schemes.  

\begin{widetext}
    \begin{figure}
        \centering
        \includegraphics[width =\textwidth]{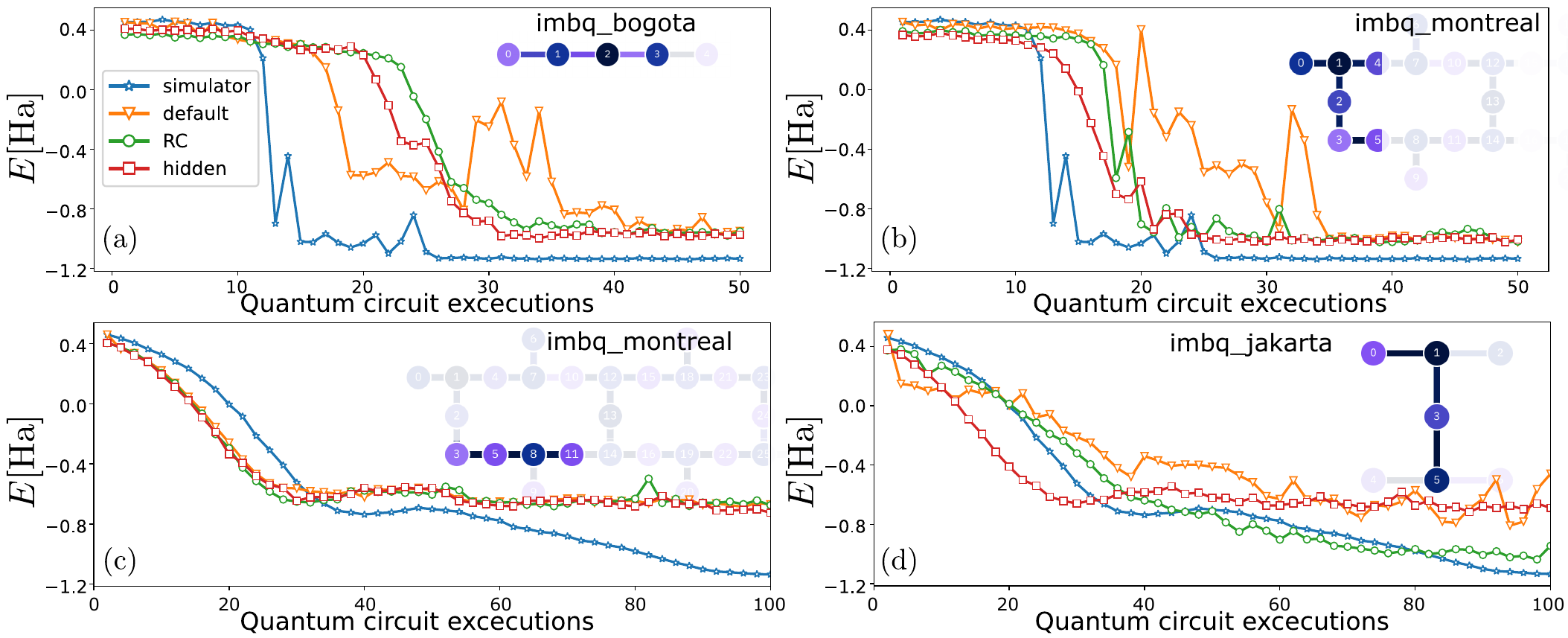}
        \caption{{\it Energy ground state estimation for $H_2$ using VQE:} In panel (a) using IBMQ-Bogota and (b) IBMQ-Montreal, we present the learning paths with BOBYQA as the optimization algorithm to update the quantum circuit parameters. We present results from randomized compilation (RC), hidden inverse (hidden), raw hardware data without QEM (default), and Aer-simulator (simulator) in each panel. In panels (c) and (d) we use the gradient-based solver Adam. Notably, coherent noise mitigation schemes showed improvement in BOBYQA convergence time, with slightly better performance from the HI circuits, while RC provided higher performance in one case using Adam. In all the experiments, we used 5000 shots. For the randomized compilation, we used 20 random circuit instances.}
        \label{fig:VQELearning}
    \end{figure}
\end{widetext}

In the experimental implementation, we consider different backends, IBMQ-Bogota, Montreal, and Jakarta, to explore coherent hardware with different noise rates. In addition, we consider low and high-fidelity sectors where the mitigation could be more or less effective. 
We evaluate HI performance by comparing the VQE learning path with an unmitigated ansatz and a randomly-compiled ansatz with a samples drawn randomly from a set of 20 circuits. 
We compare both cases to a simulated experiment using the Aer-qiskit simulator. Each experiment used 5000 shots. In Fig. \ref{fig:VQELearning}, we present the learning paths for the above setups.    

\section{Discussion} \label{sec:dis}

{\it Verification:} The real parts of the default and the inverse gates' $\chi $ matrices effectively represent the same quantum process to within measurement error.  
The differences in the imaginary parts, while statistically significant, vary widely from shot to shot, making it difficult to attribute any differences to processes versus background noise. 
From the experimental point of view, good agreement in the real part of the $\chi$-matrix appears to be enough to provide good results in the unitary folding test.

{\it Unitary folding test:} We have verified the operational equivalence between the $CX$ and $CX^{-1}$ with the profile obtained in our procedure. We consider a unitary folding test that consists of the fidelity study of the $CX$ gate when we apply a unitary ${\cal U}^n$ for different powers $n$. As we can expect, the fidelity of the resulting operation $CX {\cal U}^n$ drops as $n$ grows. In this version of the unitary folding test, the procedure can amplify systematic and coherent noise. Consider the simple coherent noise model
\begin{equation}
    CX \rightarrow  CX \cdot \mathcal{E}  \equiv G_{CX},
\end{equation}
$G_{CX}$ is the {\it on-hardware} operation and $\mathcal{E}$ the error map encloses all the $CX$ imperfections. This model shows the difference between the two folding experiments, i.e., ${\cal U}_0 = G_{CX} G_{CX} = (CX\cdot {\cal E})^2$ and ${\cal U}_1 = G_{CX} G^{-1}_{CX} = \mathbb{I}$. 
Thus, ${\cal U}_0$ is the only configuration that boosts the effects of the imperfections from the coherent noise. 
Therefore, the significant fidelity decay present in the ${\cal U}_1$ folding data (see Fig.~\ref{fig:foldingExp}(a) and (b)), suggests an important level of incoherent noise was still present in two relevant sectors of IBMQ-Montreal: qubits 0,1 and qubits 3,5, at the time this experiment was performed. 
Qubit pair 7,10 had the worst reported entangling operation fidelity. As we can see in Fig. \ref{fig:foldingExp}(c), the fidelity fluctuates for  ${\cal U}_0$ and ${\cal U}_1$ folding with no clear evidence of coherence error mitigation for any power $n$. A possible scenario that explains the the inconclusive nature of the unitary folding test on these qubits is that the HI in this case was applied to a non-self-inverse quantum operation. That is, the CX gate between these two qubits could depart enough from the ideal CX such that our assumed noise model is incorrect; and further, if the default gate is not self-inverse, we cannot expect the HI to mitigate coherent noise because cancellation will not occur.

{\it Application to VQE:} The performance of the VQE algorithm, like any other quantum-classical variational algorithm, depends partly on the selection of the classical optimizer and the circuit ansatz. We applied HI to the $CX$ gates used in the UCC3 ansatz, and observed improved convergence performance in the learning process for BOBYQA, a gradient-free solver (see Fig. \ref{fig:VQELearning} (a) and (b)) in terms of the number of iterations required to reach convergence. In Fig.~\ref{fig:VQELearning} (a) convergence is reached after 15 iterations vs 50 iterations without applying the HI. When using the Adam optimizer we observe the same performance between HI and RC on Montreal. On Jakarta, while the HI data converge faster, the RC data converges to a better minimum energy after 100 iterations. The Adam solver is an algorithm for first-order gradient-based optimization with stochastic objective functions. The algorithm is appropriate for non-stationary and very noisy gradients. Therefore, up to certain level of noise, the Adam solver is better adapted to the problem. Thus, the choice of optimal QEM method depends on ansatz, optimizer, and even the underlying qubits. However, practical considerations also play a role. In general, the application of HI should depend on the magnitude of the rotation in between CX gates; for large rotations, the inverse will amplify rather than cancel coherent noise. In the experiments shown in Fig.~\ref{fig:VQELearning}, the nature of batching and queue priority prevented a practical implementation of a conditional rotation check for HI application because it requires a circuit recompilation on a per-shot basis. Therefore the improvements shown for HI in these experiments may not be optimal for all iterations.

\begin{widetext}
    \begin{figure}
        \centering
        \includegraphics[width =0.8\textwidth]{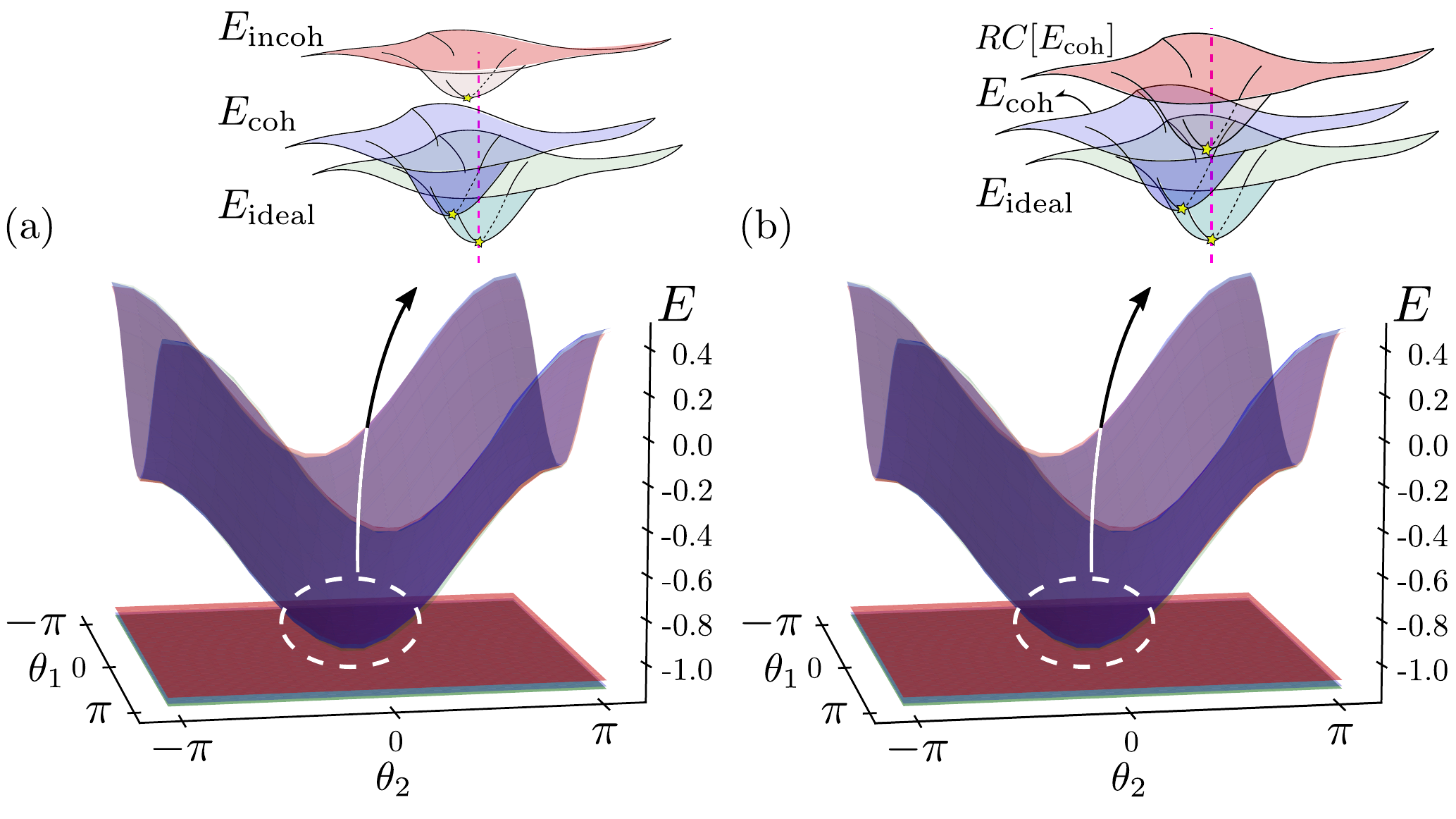}
        \caption{{\it Comparing the effect of simulated noise in VQE and the result of applying randomized compiling:} We show how a coherent or incoherent error channel modifies the loss landscape of the UCC3 ansatz for the H$_2$ molecule. (a) Loss landscape for three cases: ideal (green), subjected to coherent noise (blue), and incoherent noise (red). For each case, we draw individual planes representing the minimum of the landscape at the energy levels -1.128 Ha (ideal), -1.1256 Ha (coherent), and -1.098 Ha (incoherent). Inset displays illustration of the landscape behavior (not to scale): coherent error causes horizontal shifts of the landscape and a slight increase in the minimum energy point, while incoherent error results in flattened landscapes with increased minimum energy point. In (b), we present an identical plot as (a), with incoherent error landscapes replaced with the result of applying randomized compiling (red) to the coherent noise channel. We find RC centers the horizontal shifts but raises the minimum energy level (-1.094 Ha).}
        \label{fig:RCNoise}
    \end{figure}
\end{widetext}

To understand the effect of various noise sources and their interaction with the HI during VQE execution, we performed numerical simulations. Specifically, we construct loss landscapes of the VQE example considered above, as loss landscapes can provide key insights into the optimization problem independent of the chosen optimizer. Understanding how noise affects a loss landscape can help us make general predictions about higher level algorithm performance, particularly during optimization. In order to make three-dimensional visualization possible, we fix one of the parameters of the UCC3 ansatz and plot the calculated energy as a function of the two other free parameters. 
We consider two different noise models in these simulations. 
First, we consider a time-independent mixed unitary noise channel parameterized by $\epsilon$ (noise strength) and $\kappa$ (unitarity). 
This channel is expanded as 
\begin{equation}
   \begin{aligned} \varepsilon_{G}(\rho)=\kappa \cdot \varepsilon_{G}^{c}(\rho)+(1-\kappa) \cdot \varepsilon_{G}^{s}(\rho) \ , \ {\rm with}\\
   \end{aligned}
\end{equation}
\begin{equation}
   \begin{aligned}
     \varepsilon_{G}^{c}(\rho) &=\exp (-i \epsilon G) \rho \exp (i \epsilon G) \ {\rm and}\\ \varepsilon_{G}^{s}(\rho) &=\cos ^{2}(\epsilon) I \rho I+\sin ^{2}(\epsilon) G \rho G \ . \end{aligned}
\end{equation}
$G$ is the ideal intended operation, and $\varepsilon_{G}$ is the noisy operation with $\varepsilon_{G}^{c}$ as the coherent part and  $\varepsilon_{G}^{s}$ as the incoherent part. 
One nice feature of parameterizing the noise channel this way is that the channel fidelity is independent of $\kappa$. 
This feature allows us to generate comparable coherent and incoherent noise channels by varying $\kappa$. 
Figure \ref{fig:RCNoise}(a) compares the ideal landscape to that of a  coherent error channel ($\kappa = 1$ and $\epsilon$ = 0.02) and an incoherent error channel ($\kappa = 0$ and $\epsilon = 0.02$). 
We find that coherent error leads to horizontal shifts in the landscape compared with the ideal case and a slight increase in the minimum energy value.

\begin{widetext}
    \begin{figure}
        \centering
        \includegraphics[width =0.8\textwidth]{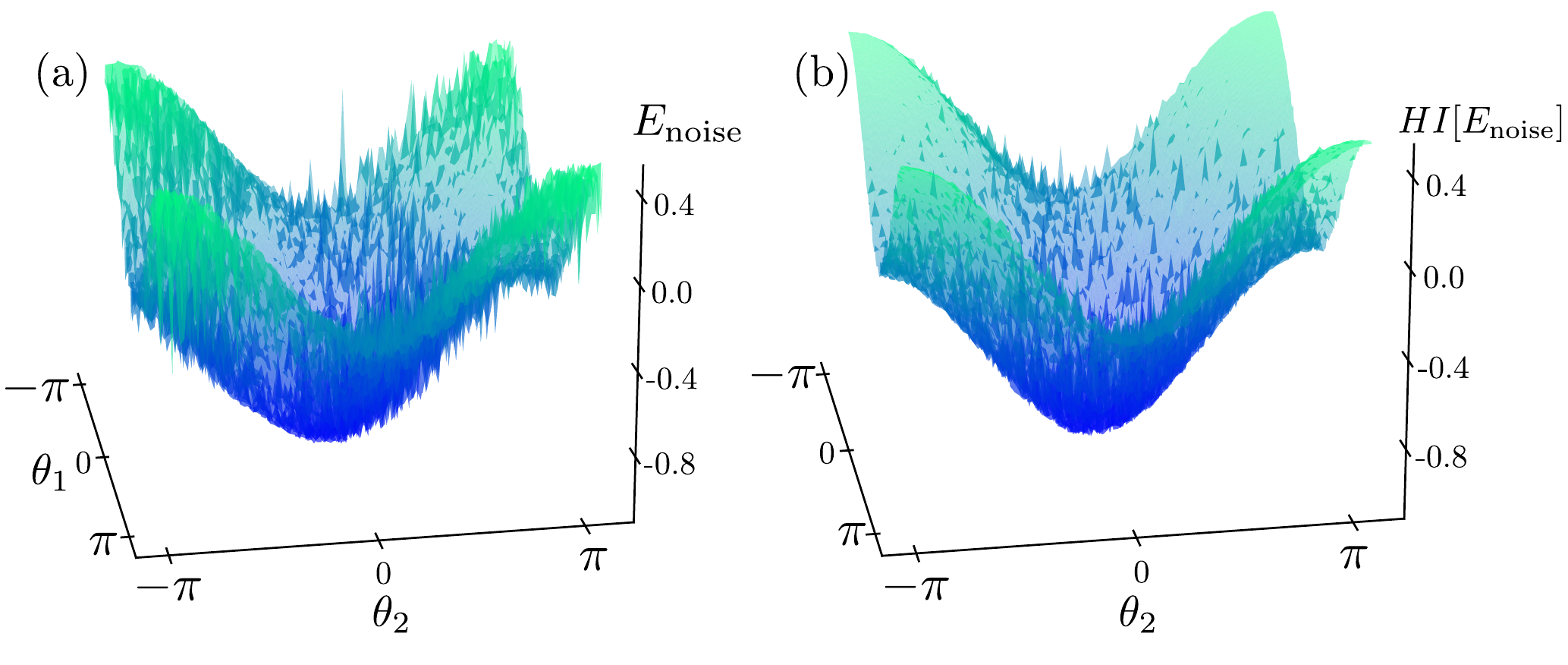}
        \caption{{\it Simulated time-dependent coherent noise in VQE and effect of HI:} Panel (a) shows the structure of the noisy loss landscape, and panel (b), shows the same landscape with HI applied.}
        \label{fig:HINoise}
    \end{figure}
\end{widetext}

On the other hand, incoherent error leads to flatter landscapes with a noticeable difference in the minimum energy value. 
As a result, one can expect a better eigenvalue approximation from coherent errors and a better eigenvector approximation from incoherent errors of similar noise strengths. 
We also simulated the effect of randomized compiling in the presence of coherent errors, as seen in figure \ref{fig:RCNoise}(b). 
We find randomized compiling aims to re-center the horizontally shifted landscapes (due to coherent errors) but increases the minimum energy value.  
In addition, we consider the action of the HI on an induced coherent noise by over-rotation and show how this affects the loss landscape (see Figs. \ref{fig:HINoise}(a) noisy, and (b) mitigated). 
We consider a more damaging time-dependent coherent error in the two-qubit entangling operations with $[ZX]_{(1+\epsilon)\theta}$ with $\epsilon \sim  \mathcal{N}(0, \varepsilon)$. 
We calculated a root mean square surface roughness of the loss landscape plot as 0.1065 for the default Ansatz and 0.0611 for the HI-optimized Ansatz. 
We observe a partial recovery of the loss landscape from the noise influence by applying the hidden-inverse. 
 
This result gives a reasonable inference on the possible reasons for the energy converged values in the on-hardware experiments. 
Due to incoherent noise, the hidden-inverse and the RC do not improve the converged energy value in the VQE learning path, while they do improve convergence. 
Applying these methods in combination with another QEM technique, such as readout error mitigation, would potentially yield both faster convergence and a better minimum energy.
 
\section{Conclusions} \label{sec:conc}
In this work, we studied the effects of inverse gates on the performance of a variational quantum algorithm and compared it with randomized compiling. The HI implementation does not require prior knowledge of the noise source aside from the coherent noise model, and does not require extra quantum resources (gates). We depart from the default pulse structure of the gates to build their inverses and apply them to one the latter gate of each pair of CX's in the circuit to enable noise cancelation. The mitigation provides a boost to the performance of BOBYQA, reducing the number of iterations, and thus quantum resources, required to reach a converged value. We further explored the effectiveness of randomized compiling and noticed its capability of reducing coherent noise as well, at the cost of additional single qubit gates at each CX location in the circuit. Simulations suggest that incoherent noise is also present in the on-hardware experiments, which also matches with prior experience and VQE experiments on superconducting platforms, which explains the higher ground state found in several cases regardless of the coherent error mitigation method used.  
The results show that the HI mitigation mechanism, previously demonstrated on trapped ion platforms, is also effective on superconducting hardware. Combining this QEM method with incoherent noise mitigation schemes should yield high improved performance in future VQE applications. The HI gate construction technique is available as open source and in a pending qiskit pull request.

\begin{acknowledgments}
Authors thank Kenneth Brown for helpful discussions. This work was supported as part of the ASCR Quantum Testbed Pathfinder Program at Oak Ridge National Laboratory under FWP \# ERKJ332.  S.M. was supported through US Department of Energy grant DE-SC0019294 awarded to Duke and is funded in part by an NSF QISE-NET fellowship (1747426). This research used resources of the Oak Ridge Leadership Computing Facility, which is a DOE Office of Science User Facility supported under Contract DE-AC05-00OR22725.
\end{acknowledgments}
\section*{Author contributions}
    V.L.O. performed on-hardware experiments, S.M. performed simulations, R.C.P.  ~analyzed and interpreted experimental data and suggested experiments to run. V.L.O, S.M., and R.C.P. interpreted the results and co-wrote the manuscript.
\section*{Competing interests}
    The authors declare that there are no competing interests.

\appendix

\section{$\pi$-pulse on superconducting devices}
We transform the default pulse structure of a given quantum operation $G$ to create its inverse $G^{-1}$. We can illustrate the inversion by a simple example, a driven two-level system with the Hamiltonian model ($\hbar = 1$ throughout)
\begin{equation} \label{eq:tls}
H_{\rm TLS} (t) = \omega_0 \sigma^z/2 + u(t) \varepsilon \cos(\omega_1 t) \sigma^x ,
\end{equation}
with $\omega_0$ as the system's proper energy, $\omega_1$ as the external driving modulation, $\sigma^z$ and $\sigma^x$ as Pauli matrices, and $\varepsilon$ as the external driving strength.  We have considered a profile term $u(t)$ for the external driving that controls the dynamics. Under a weak driving strength regime, where $\varepsilon \ll \omega_0$, and around the resonance condition $\omega_1 \approx \omega_0$, we can invoke the rotating wave approximation (RWA) and describe the coherent dynamics with the Hamiltonian 
\begin{equation}
 H^{\rm RWA}_{\rm TLS}(t) \approx u(t) \varepsilon \sigma^x /2 \ .  
\end{equation}
We have considered $u(t)$ a slow function for external driving, therefore we can keep that term in the RWA final expression. Now, the quantum process can be determined by the evolution operator in this picture
\begin{equation}
 G(T) = \exp \left[   - i    \int_0^T dt \, H^{\rm RWA}_{\rm TLS}(t) \right]=\exp \left[   - i    \sigma^x \, \theta(T)/2 \right] , 
\end{equation}
with $\theta(T) = \varepsilon \int_0^T dt \, u(t) $ as the accumulated phase by the pulse profile $u(t)$ in the interval of time $[0,T]$. In order to drop the quantum process time dependence, we can consider $u(t)$ finite in the interval  $[0,T]$ and negligible or zero in the rest of the domain. The Gaussian profile is a common choice for the pulse, i.e., $u(t) = u_0 \exp[- (t - T/2)^2/(2 u_\sigma^2)]$, with amplitude $u_0$, width $u_\sigma$, and centered at $T/2$. With this choice, the accumulated phase becomes $\theta(T) = \varepsilon u_0 u_\sigma \sqrt{2 \pi }$, therefore, the phase and the quantum process $G$ can be controlled by the amplitude $u_0$ and the width $u_\sigma$. For $\theta(T) = \pi$, we can generate the unique case $G = \sigma_x \equiv X$, and the profile $u(t)$ that follows that structure is called $\pi$-pulse. 

In a general case, where $u(t)$ does not follow a specific function, the profile can be described by a  piece-wise function,
\begin{equation}
 u(t) =  \left\lbrace \begin{array}{ccc}
  u_k & {\rm for} & k \delta t < t \leq (k+1) \delta t, \ k = 0, \cdots, N  \\
  0 & & {\rm otherwise}
\end{array} 
\right. .
\end{equation}
Thus, the profile is determined for the set of amplitudes $\lbrace u_0, \cdots, u_N \rbrace$, the time $\delta t$, and the number of blocks $N$. With this profile structure, the quantum process can be written as 
\begin{equation}
 G = \Pi_{k = 0}^{N} \exp [-i \varepsilon u_k \delta t \ \sigma^x/2] . 
\end{equation}
 Therefore, for the inversion of the quantum process $G$ we need to change the set of amplitudes $\lbrace u_0, \cdots , u_N \rbrace$ to the new set $\lbrace -u_N, -u_{N-1}, \cdots, -u_0 \rbrace$. We extract the amplitudes for each channel in the default pulse structure of a given gate $G$, and transform that information to generate a new pulse structure that represents the inverse $G^{-1}$.

\section*{References}
\bibliographystyle{iopart-num}
\bibliography{biblio}

\end{document}